\documentclass[prd,showpacs,preprintnumbers,floatfix]{revtex4}
\usepackage{amsmath}
\begin{document}
\title{The Casimir effect for a massless spin-3/2 field in Minkowski spacetime}
\date{\today }
\author{Wenbiao Liu}
\email{wbliu@bnu.edu.cn}
\affiliation{Department of Physics,
Beijing Normal University, Beijing, 100875, China}
\author{Kui Xiao}
\email{87xiaokui@163.com} \affiliation{Department of Physics,
Beijing Normal University, Beijing, 100875, China}
\author{Hongbao Zhang}
    \email{hbzhang@pkuaa.edu.cn}
    \affiliation{Department of Astronomy, Beijing Normal University, Beijing, 100875,
    China\\
    Department of Physics,
Beijing Normal University, Beijing, 100875, China\\
 CCAST (World
Laboratory), P.O. Box 8730, Beijing,
   100080, China}
\begin{abstract}
The Casimir effect has been studied for various quantum fields in
both flat and curved spacetimes. As a further step along this
line, we provide an explicit derivation of Casimir effect for
massless spin-3/2 field with periodic boundary condition imposed
in four-dimensional Minkowski spacetime. The corresponding results
with Dirichlet and Neumann boundary conditions are also discussed.
\end{abstract}

\maketitle

\section{Introduction}
Since the pioneering work by Casimir in 1948, who theoretically
predicted that there is an attractive force between a pair of
neutral plane metallic plates due to a shift in the energy of
vacuum state of the quantized electromagnetic field\cite{Casimir},
along with the first experimental confirmation by Sparnaay in
1958\cite{Sparnaay}, the so called Casimir effect has been studied
both extensively and intensively. Not only has the Casimir effect
been measured more and more precisely in recent experiments
\cite{Lamoreauxb,MR,Roy,Ederth,Chana,Chanb,Bressi,Iannuzzi}, but
also acquired interesting applications in different areas of
theoretical physics, including the QCD bag model of
hadrons\cite{bender,Miltonn, Miltonm} and dark energy in
cosmology\cite{Miltonl}. For more recent progress and
developments, please refer to \cite{Bordag,Milton04,Lamoreauxa}
and references therein.

On the other hand, by many methods such as Green's
function\cite{Lowell}, zeta function\cite{Hawking}, path
integral\cite{symanzik}, dimensional regularization\cite{Treml}, and
cut-off method\cite{svaiter,Edery}, the Casimir effect has been
computed for various fields which satisfy different boundary
conditions in both flat and curved
spacetimes\cite{Ford,Ambjorn,Miltona,KMiltona,KMiltonb,Birmingham,MiltonJa,
MiltonJb,Bayin,Langfeld,Aliev,Miltonb,Elizalde,SST,CEK,MF,EST,Setare,SS,SJaffe,SB,BPT}.
However, as far as a spin-3/2 field is concerned, we are not aware
of any work done in this direction. Therefore, as a further step
along this line, the present paper will provide an explicit
derivation of the Casimir effect for a massless spin-3/2 field with
the periodic boundary condition in four-dimensional Minkowski
spacetime.

It is obvious that the massless spin-3/2 field occupies a special
position in our attempts to understand nature both relativistically
and quantum mechanically. It is the massless spin-3/2 field that
turns out to be the simplest nontrivial higher spin field, but plays
a significant role in supergravity and twistor programme.
Especially, different from neutrino and electromagnetic fields,
there is no gauge invariant local energy momentum tensor for
massless spin-3/2 field, which is also shared by linear
gravitational field indeed. However, fortunately, as is shown in
\cite{SZ}, the integral of energy momentum tensor over the whole
space is gauge independent, which justifies our present work and
makes it acquire particular interest.

The paper is organized as follows. In the next section, we will
briefly review the gauge invariant theory of massless spin-3/2 field
in four-dimensional Minkowski spacetime. In the subsequent section,
introducing zeta function, we provide an explicit derivation of the
Casimir effect for massless spin-3/2 field with periodic boundary
condition, which means that we compactify the corresponding space to
a circle. Conclusions and discussions are given in the final
section.

System of natural unites are adopted: $\hbar=c=1$. Notations and
conventions follow those in \cite{SZ}. Especially, the metric
signature takes $(+,-,-,-)$, and
$\{\sigma^\mu{_{\Sigma'\Sigma}}=\frac{1}{\sqrt{2}}(I,\sigma)|\mu=
0,1,2,3;\Sigma(\Sigma')=1,2\}$ with $\sigma$ Pauli matrices.
\section{Equation of motion and energy momentum tensor for massless spin-3/2
field from Rarita-Schwinger Lagrangian}

This section will present a brief review of the theory of massless
spin-3/2 field in four-dimensional Minkowski spacetime, which
provides a concise foundation for later work. For more details,
please refer to \cite{SZ}.

Start with Rarita-Schwinger Lagrangian\cite{SZ,RS}
\begin{equation}
\mathcal{L}=-i\sqrt{2}[\bar{\psi}^{aB^{\prime }}\sigma^b{_{B^{\prime }B}}%
\nabla_b\psi_a{^B}-\frac{1}{3}(\bar{\psi}^{aB^{\prime }}\sigma_{aB^{\prime
}B}\nabla_b\psi^{bB}+\bar{\psi}^{aB^{\prime }}\sigma_{bB^{\prime
}B}\nabla_a\psi^{bB})+\frac{2}{3}\bar{\psi}^{aB^{\prime }}\sigma_{aB^{\prime
}B}\sigma^{bBC^{\prime }}\sigma_{cC^{\prime }C}\nabla_b\psi^{cC}],
\label{Lagrangian}
\end{equation}
where the bar denotes the Hermitian conjugation. From here, Euler-Lagrange
equation leads to
\begin{equation}
\sigma^b{_{B^{\prime }B}}\nabla_b\psi_a{^B}-\frac{1}{3}(\sigma_{aB^{\prime
}B}\nabla_b\psi^{bB}+\sigma_{bB^{\prime }B}\nabla_a\psi^{bB})+\frac{2}{3}%
\sigma_{aB^{\prime }B}\sigma^{bBC^{\prime }}\sigma_{cC^{\prime
}C}\nabla_b\psi^{cC}=0.
\end{equation}
With the covariant derivative and the soldering form action on the equation
of motion, respectively, we have
\begin{eqnarray}
\sigma^b{_{B^{\prime }B}}\nabla_b\nabla^a\psi_a{^B}&=&0,  \notag \\
\nabla^a\psi_a{^B}&=&0,
\end{eqnarray}
where the identity $\sigma_{aCB^{\prime }}\sigma_{bD}{^{B^{\prime }}}%
+\sigma_{bCB^{\prime }}\sigma_{aD}{^{B^{\prime }}}=\eta_{ab}\epsilon_{CD}$
has been employed\cite{SZ}. Taking into account Rarita-Schwinger constraint
condition, i.e.,
\begin{equation}
\sigma^a{_{B^{\prime }B}}\psi_a{^B}=0,  \label{RS1}
\end{equation}
the equation of motion is simplified as
\begin{equation}
\sigma^b{_{B^{\prime }B}}\nabla_b\psi_a{^B}=0.  \label{RS2}
\end{equation}
Eqn.(\ref{RS1}) and Eqn.(\ref{RS2}) are just our familiar
Rarita-Schwinger equations for massless spin-3/2
field\cite{SZ,RS}. Furthermore, by Belinfante's construction and
after a straightforward calculation, the local energy momentum
tensor for massless spin-3/2 field reads\cite{SZ}
\begin{equation}
T^{ab}=-i\sqrt{2}[\frac{1}{2}(\bar{\psi}^{dD^{\prime
}}\sigma^{(b}{_{D^{\prime }E}}\nabla^{a)}\psi_d{^E}-\nabla^{(a}\bar{\psi}%
^{|dD^{\prime }|}\sigma^{b)}{_{D^{\prime }E}}\psi_d{^E})+(\nabla_c\bar{\psi}%
^{(b|D^{\prime }|}\sigma^{a)}{_{D^{\prime }D}}\psi^{cD}-\bar{\psi}%
^{cD^{\prime }}\sigma^{(a}{_{D^{\prime }D}}\nabla_c\psi^{b)D})],  \label{em}
\end{equation}
which is equivalent with that obtained by the variational principle\cite%
{Zhang}.

It is worth noting that Rarita-Schwinger field equations are invariant under
the following gauge transformation\cite{SZ,RS}
\begin{equation}
\psi_a{^B}\rightarrow\psi_a{^B}+\nabla_a\varphi^B
\end{equation}
with
\begin{equation}
\sigma^b{_{B^{\prime }B}}\nabla_b\varphi^B=0.
\end{equation}
Moreover, the global energy is also gauge invariant, although the
local energy momentum tensor (\ref{em}) including the energy
density is gauge dependent\cite{SZ}. Since the Casimir effect
involves the global energy rather than local energy density, in
the following discussions we can confine ourselves to Coulomb
gauge, i.e.,
\begin{equation}
\psi_0{^B}=0.
\end{equation}
Obviously, the energy density in Coulomb gauge is given by
\begin{equation}
\rho=T^{00}=-i\frac{\sqrt{2}}{2}(\bar{\psi}^{dD^{\prime }}\sigma^0{%
_{D^{\prime }E}}\nabla^0\psi_d{^E}-\nabla^0\bar{\psi}^{dD^{\prime }}\sigma^0{%
_{D^{\prime }E}}\psi_d{^E}).  \label{density}
\end{equation}

For later progress in the subsequent section, a consistent
massless spin-3/2 quantum field can be constructed by the plane
wave basis in Coulomb gauge as\cite{SZ}
\begin{equation}
\hat{\psi}_{a}{^{B}}(x)=\frac{1}{\sqrt{(2\pi )^{3}}}\int d^{3}\mathbf{p}[a(\mathbf{p})\psi _{p}{_{a}{^{B}%
}}(x)+c^{\dag }(\mathbf{p})\psi _{-p}{_{a}{^{B}}}(x)],p_{0}>0.
\label{quantum}
\end{equation}%
Here the annihilation and creation operators satisfy the anti-commutation
relations as follows
\begin{eqnarray}
\{a(\mathbf{p}),a(\mathbf{p^{\prime }})\} &=&0,  \notag \\
\{a(\mathbf{p}),a^{\dag }(\mathbf{p^{\prime }})\} &=&\delta ^{3}(\mathbf{p}-%
\mathbf{p^{\prime }}),  \notag \\
\{a^{\dag }(\mathbf{p}),a^{\dag }(\mathbf{p^{\prime }})\} &=&0,  \notag \\
\{c(\mathbf{p}),c(\mathbf{p^{\prime }})\} &=&0,  \notag \\
\{c(\mathbf{p}),c^{\dag }(\mathbf{p^{\prime }})\} &=&\delta ^{3}(\mathbf{p}-%
\mathbf{p^{\prime }}),  \notag \\
\{c^{\dag }(\mathbf{p}),c^{\dag }(\mathbf{p^{\prime }})\} &=&0.
\label{anti-commutation}
\end{eqnarray}%
The plane wave solutions to Rarita-Schwinger equations in Coulomb gauge read
\begin{equation}
\psi _{p}{_{a}{^{B}}}(x)=\frac{1}{\sqrt{2|p_{0}|}%
}\tilde{\psi}_{\mu }{^{\Sigma }}(p)(dx^{\mu })_{a}(\varepsilon
_{\Sigma })^{B}e^{-ip_{b}x^{b}},
\end{equation}%
where
\begin{equation}
\tilde{\psi}(1,0,0,1)=(0,1,i,0)\otimes \left(
\begin{array}{ll}
1 &  \\
0 &
\end{array}%
\right) ,
\end{equation}%
and
\begin{eqnarray}
\tilde{\psi}_{\mu }{^{\Sigma }}(p=e^{-\lambda },e^{-\lambda }\sin \theta
\cos \varphi ,e^{-\lambda }\sin \theta \sin \varphi ,e^{-\lambda }\cos
\theta ) &=&\tilde{\psi}_{\mu }{^{\Sigma }}(-p)  \notag \\
&=&(\Lambda ^{-1})^{\nu }{_{\mu }}L^{\Sigma }{_{\Gamma }}\tilde{\psi}_{\nu }{%
^{\Gamma }}(1,0,0,1)
\end{eqnarray}%
with
\begin{eqnarray}
\Lambda &=&\left(
\begin{array}{cccc}
1 & 0 & 0 & 0 \\
0 & \cos \varphi & -\sin \varphi & 0 \\
0 & \sin \varphi & \cos \varphi & 0 \\
0 & 0 & 0 & 1%
\end{array}%
\right) \left(
\begin{array}{cccc}
1 & 0 & 0 & 0 \\
0 & \cos \theta & 0 & \sin \theta \\
0 & 0 & 1 & 0 \\
0 & -\sin \theta & 0 & \cos \theta%
\end{array}%
\right) \left(
\begin{array}{cccc}
\cosh \lambda & 0 & 0 & -\sinh \lambda \\
0 & 1 & 0 & 0 \\
0 & 0 & 1 & 0 \\
-\sinh \lambda & 0 & 0 & \cosh \lambda%
\end{array}%
\right) ,  \notag \\
L &=&\left(
\begin{array}{cc}
e^{-i\frac{\varphi }{2}} & 0 \\
0 & e^{i\frac{\varphi }{2}}%
\end{array}%
\right) \left(
\begin{array}{cc}
\cos \frac{\theta }{2} & -\sin \frac{\theta }{2} \\
\sin \frac{\theta }{2} & \cos \frac{\theta }{2}%
\end{array}%
\right) \left(
\begin{array}{cc}
e^{-\frac{\lambda }{2}} & 0 \\
0 & e^{\frac{\lambda }{2}}%
\end{array}%
\right) .
\end{eqnarray}
\section{The Casimir effect for massless spin-3/2 field with periodic boundary condition}
Now consider the constraint of having a rectangular box with the
boundaries located in
$x_1=-\frac{L_x}{2},y_1=-\frac{L_y}{2},z_1=0$ and
$x_2=\frac{L_x}{2},y_2=\frac{L_y}{2},z_2=L_z$ respectively. Thus
we can implement the periodic boundary condition on the massless
spin-3/2 quantum field, i.e.,
\begin{eqnarray}
\hat{\psi}_{\mu }{^{\Sigma }}(x_1,y,z)&=&\hat{\psi}_{\mu }{
^{\Sigma }}(x_2,y, z), \nonumber\\
\hat{\psi}_{\mu }{^{\Sigma }}(x,y_1,z)&=&\hat{\psi}_{\mu }{
^{\Sigma }}(x,y_2, z), \nonumber\\
\hat{\psi}_{\mu }{^{\Sigma }}(x,y,z_1)&=&\hat{\psi}_{\mu }{
^{\Sigma }}(x,y, z_2),
\end{eqnarray}
hence the expression (\ref{quantum}) can be modified to
\begin{equation}
\hat{\psi}_{a}{^{B}}(x)=\frac{1}{\sqrt{L_xL_yL_z}}\{\sum_{p_x=\frac{2m\pi}{L_x}}\sum_{p_y=\frac{2n\pi}{L_y}}\sum_{p_z=\frac{2l\pi}{L_z}}[a(\mathbf{p})\psi
_{p}{_{a}{^{B}}}(x)+c^{\dag }(\mathbf{p})\psi
_{-p}{_{a}{^{B}}}(x)]\},p_{0}>0\label{restriction}
\end{equation}
with
\begin{eqnarray}
\{a(\mathbf{p}),a(\mathbf{p^{\prime }})\} &=&0,  \nonumber \\
\{a(\mathbf{p}),a^{\dag }(\mathbf{p^{\prime }})\} &=&\delta_{p_xp_x'}\delta_{p_yp_y'}\delta_{p_zp_z'},  \nonumber \\
\{a^{\dag }(\mathbf{p}),a^{\dag }(\mathbf{p^{\prime }})\} &=&0,  \nonumber \\
\{c(\mathbf{p}),c(\mathbf{p^{\prime }})\} &=&0,  \nonumber \\
\{c(\mathbf{p}),c^{\dag }(\mathbf{p^{\prime }})\} &=&\delta_{p_xp_x'}\delta_{p_yp_y'}\delta_{p_zp_z'},  \nonumber \\
\{c^{\dag }(\mathbf{p}),c^{\dag }(\mathbf{p^{\prime }})\} &=&0.
\label{relation restriction}
\end{eqnarray}

Next substituting the  massless spin-3/2 quantum field
(\ref{restriction}) into Eqn.(\ref{density}) and taking the integral
over the box, the expectation value of the Casimir energy in the
quantum vacuum state can be written as
\begin{eqnarray}
E&=&\int dxdydz\langle0|\hat{\rho}|0\rangle =\int dxdydz\langle0|\hat{T}^{00}|0\rangle  \nonumber \\
&=&\frac{1}{2}\frac{1}{L_xL_yL_z}\int
dxdydz\Big\{\sum_\mathbf{p}\sum_\mathbf{p'}
\frac{1}{\sqrt{2p_{0}}} \frac{1}{\sqrt{2p'_{0}}}(-\sqrt{2})
\nonumber
\\
&&\{(p_{0}+p'_{0})[\langle0| a^{\dag
}(\mathbf{p})a(\mathbf{p'})|0\rangle \bar{\tilde{\psi}}^{\mu
\Sigma'}(p)\sigma ^{0}{_{\Sigma'\Sigma
}}\tilde{\psi}_{\mu }{^{\Sigma }}(p')e^{i(p_{b}-p'_{b})x^{b}}  \nonumber \\
&&-\langle0| c(\mathbf{p})c^{\dag }(\mathbf{p'})|0\rangle
\bar{\tilde{ \psi}}^{\mu \Sigma'}(-p)\sigma ^{0}{_{\Sigma'\Sigma
}} \tilde{\psi}_{\mu }{^{\Sigma }}(-p')e^{-i(p_{b}-p'_{b}
)x^{b}}]  \nonumber \\
&&+(p_{0}-p'_{0})[\langle0| a^{\dag }(\mathbf{p})c^{\dag
}(\mathbf{ p'})|0\rangle \bar{\tilde{\psi}}^{\mu \Sigma'
}(p)\sigma ^{0} {_{\Sigma'\Sigma }}\tilde{\psi}_{\mu
}{^{\Sigma }}(-p')e^{i(p_{b}+p'_{b})x^{b}}  \nonumber\\
&&-\langle0| c(\mathbf{p})a(\mathbf{p'})|0\rangle
\bar{\tilde{\psi}} ^{\mu \Sigma'}(-p)\sigma ^{0}{_{\Sigma'\Sigma
}}\tilde{ \psi}_{\mu }{^{\Sigma
}}(p')e^{-i(p_{b}+p'_{b})x^{b}}]\}\Big\}.\label{Casimir}
\end{eqnarray}
By the anti-commutation relations (\ref{relation restriction}) as
well as the following properties of annihilation and creation
operator, i.e.,
\begin{eqnarray}
a(\mathbf{p})|0\rangle=0,\nonumber\\
c(\mathbf{p})|0\rangle=0,\nonumber\\
\langle0|a^\dag(\mathbf{p})=0,\nonumber\\
\langle0|c^\dag(\mathbf{p})=0,
\end{eqnarray}
Eqn.(\ref{Casimir}) can be simplified as
\begin{eqnarray}
E&=&\frac{1}{2}\sum_\mathbf{p}[\sqrt{2}\bar{\tilde{ \psi}}^{\mu
\Sigma'}(-p)\sigma ^{0}{_{\Sigma'\Sigma }} \tilde{\psi}_{\mu
}{^{\Sigma }}(-p)]\nonumber\\
&=&\frac{1}{2}\sum_\mathbf{p}[\sqrt{2}\Lambda ^{0}{%
_{\nu }}\bar{\tilde{\psi}}^{\mu \Sigma'}(1,0,0,1)\sigma ^{\nu }{%
_{\Sigma'\Sigma }}\tilde{\psi}_{\mu }{^{\Sigma
}}(1,0,0,1)]\nonumber\\
&=&\frac{1}{2}\sum_\mathbf{p}(-2p_{0})=-\sum_\mathbf{p}p_0.
\end{eqnarray}
In order to investigate the Casimir effect between two neutral
infinite parallel plane plates without loss of generalization, we
shall simultaneously put $L_x\rightarrow\infty$ and
$L_y\rightarrow\infty$ at the end of our calculation. Thus an
integral approximation for the sum over the modes of $p_x$ and
$p_y$ is possible, i.e.,
\begin{eqnarray}
E&=&-\frac{L_xL_y}{(2\pi)^2}[\sum_{p_z=\frac{2l\pi}{L_z}}\int
dp_xdp_y(\sqrt{p_x^2+p_y^2+p_z^2})]\nonumber\\
&=&-\frac{L_xL_y}{(2\pi)^2}[\sum_{p_z=\frac{2l\pi}{L_z}}\int
d^2\mathbf{p}_\tau(\sqrt{\mathbf{p}_\tau^2+p_z^2})]\nonumber\\
&=&-2\frac{L_xL_y}{(2\pi)^2}[\sum_{l=1}^{\infty}\int
d^2\mathbf{p}_\tau(\sqrt{\mathbf{p}_\tau^2+(\frac{2l\pi}{L_z})^2})],\label{result}
\end{eqnarray}
where $\mathbf{p}_\tau=(p_x,p_y)$ is introduced for the
convenience of later computation, and the mode for $l=0$ is
ignored for it carries zero longitudinal momentum, thus does not
contribute to the Casimir effect.

Obviously, Eqn.(\ref{result}) is divergent. To extract a finite
value from Eqn.(\ref{result}), we here invoke zeta function
regularization. Firstly introducing the definition of the gamma
function as follows
\begin{equation}
\lambda^{-s}\Gamma(s)=\int_0^\infty \frac{dt}{t} t^s e^{-\lambda
t},
\end{equation}
then the total energy can be written as
\begin{eqnarray}
E&=&-2\frac{L_xL_y}{(2\pi)^2\Gamma({-\frac{1}{2}})}\sum_{l=1}^{\infty}\int
d^2 \mathbf{p}_{\tau}\int_0^\infty\frac{dt}{t}t^{-1/2}
e^{-t({{\mathbf{p}}_\tau}^2+l^24\pi^2/{L_z}^2)}, \label{res}
\end{eqnarray}
next interchanging the integral over $\mathbf{p}_\tau$ and $t$, and
noticing that the integral over $\mathbf{p}_\tau$ is just Gaussian
type integral, we have
\begin{eqnarray}
E &=&\frac{L_xL_y}{(4\pi)\sqrt{\pi }}\sum_{l=1}^\infty\int
\frac{dt}{t}t
^{-3/2}e^{-tl^{2}(2\pi)^{2}/L_z^{2}}  \nonumber \\
&=&\frac{L_xL_y}{4\pi\sqrt{\pi
}}(\frac{2\pi}{L_z})^{3}\Gamma(-\frac{3}{2})\zeta(-3) ,
\label{res}
\end{eqnarray}
where we have used the Riemann zeta function
\begin{equation}
\pi ^{-s/2}\zeta(s)\Gamma(\frac{s}{2})
=\sum_{l=1}^\infty\int_0^\infty \frac{dt}{t}t^{s/2} e^{-tl^{2}\pi
}.
\end{equation}
Later by $\Gamma(-{\frac{3}{2}})=\frac{4\sqrt{\pi}}{3}$, and
$\zeta({-3})=\frac{1}{120}$, we obtain
\begin{equation}
E=\frac{\pi ^{2}{L_xL_y}}{45L_z^{3}}.
\end{equation}

Furthermore, the corresponding Casimir force per unit area between
the two parallel plane plates is obtained by taking the negative
derivative of $\frac{E}{L_xL_y}$ with respect to the distance of
the two plates $L_z$, i.e.,
\begin{equation}
f=\frac{\pi ^{2}}{15L_z^{4}}.
\end{equation}
Obviously, unlike the case of scalar field, the Casimir force is
repulsive for the massless spin-3/2 field, which originates from
the spin-statistics connection: the massless spin-3/2 field
satisfies Fermi-Dirac statistics rather than Bose-Einstein
statistics. In addition, the magnitude of the resultant force is
two times of that of scalar field with the same boundary
condition, which is in agreement with the fact that the massless
spin-3/2 field has the degrees of freedom with two times as many
as the scalar one.
\section{Conclusions and Discussions}
Starting from Rarita-Schwinger Lagrangian and employing the gauge
invariance of the global energy for massless spin-3/2 field, we
have provide an explicit derivation of the Casimir effect for the
massless spin-3/2 field with periodic boundary condition in
Coulomb gauge by the zeta function regularization method. The
resultant Casimir force obtained here, together with those of
other massless fields with the same boundary condition in
four-dimensional Minkowski spacetime, can be casted in terms of
unified form as follows\cite{Ambjorn}
\begin{equation}
f=\pm\frac{S\pi ^{2}}{30L_z^{4}},
\end{equation}
where $\pm$ corresponds to fermionic and bosonic fields, implying
repulsion and attraction of the force, respectively. In addition,
$S$ denotes the degrees of freedom for fields: $S=1$ for scalar
field, and $S=2$ for massless fields with spin.

We conclude with some discussions in order. Firstly, although the
present paper only explicitly evaluates the Casimir effect for the
massless spin-3/2 field with periodic boundary condition, the
corresponding results with Dirichlet and Neumann boundary
conditions can be easy to obtain as is shown for the case of
scalar field in \cite{Ambjorn}: it is smaller, by a factor of 16,
than that with periodic boundary condition here. In addition, it
is natural to next calculate the Casimir effect for the massless
spin-3/2 field confined in other configurations such as a cylinder
and a sphere. However, under these circumstances, the calculations
involved are much more complex, thus worthy of further
investigation, which is expected to be reported elsewhere.
\section*{Acknowledgements}
W. Liu and K. Xiao are supported partly by NSFC(Grant No.10373003
and Grant No.10475013) and NBRPC(Grant No.2003CB716302). H. Zhang's
work is supported in part by NSFC(Grant No.10373003 and Grant
No.10533010).

\end{document}